\documentclass[conference]{IEEEtran}
\IEEEoverridecommandlockouts
\usepackage{cite}
\usepackage{amsmath,amssymb,amsfonts}
\usepackage{algorithmic}
\usepackage{graphicx}
\usepackage{textcomp}
\usepackage{xcolor}
\usepackage{float}
\def\BibTeX{{\rm B\kern-.05em{\sc i\kern-.025em b}\kern-.08em
    T\kern-.1667em\lower.7ex\hbox{E}\kern-.125emX}}
\begin{document}

\newcommand{\circnum}[1]{\textcircled{\scriptsize #1}}
\title{Towards Scaling Qualitative Analysis of Video Data\\
}

\author{\IEEEauthorblockN{Shiyi He}
\IEEEauthorblockA{\textit{School of Computing, University of Utah} \\
Salt Lake City, US \\
shiyi.he@utah.edu}
}

\maketitle

\begin{abstract}
Scaling qualitative video analysis is difficult as studies grow. This paper presents QualiVision, a design probe examining how an interactive, spreadsheet-backed workspace can support video-based qualitative analysis. By integrating video, transcripts, coding streams, preliminary reports, heuristic visualization and AI support, QualiVision aims to help researchers preserve evidence, compare interpretations, and conduct iterative, reflexive sensemaking as their analysis evolves.
\end{abstract}


\section{Introduction}


Video-based qualitative analysis has been widely used to analyze various interaction contexts and identify usability problems. Compared to static artifacts or retrospective accounts, video data preserve a rich resource of temporal, multimodal interactions while preserving the surrounding context, making it valuable for studying complex, situated, real-world activity.

Yet scaling qualitative analysis of video data becomes challenging as studies grow in size and complexity. Video analysis is inherently labor-intensive, often requiring 5-10 times the original playback duration~\cite{b1}. As studies expand, researchers must compare evidence across more video sessions and participants, coordinate interpretations across multiple coders, and maintain links among data, codes, and observations. What may be manageable in a small study can quickly become difficult to sustain when data and analytic records multiply.

This challenge exposes two persistent issues in current workflows. First, video-based qualitative analysis is often fragmented across tools. Researchers move between video players, transcript editors, spreadsheets, computational notebooks, and memo documents, making it difficult to keep evidence, codes, observations, and interpretations in a coherent analytic context throughout. Second, early-stage qualitative analysis is inherently uncertain and iterative. Researchers may not know in advance which events or patterns will become meaningful. Codes are created, reorganized, discarded, or stabilized as interpretations develop. Many existing workflows preserve the final coding structure rather than the analytic process that produced it, limiting support for comparing patterns, revisiting evidence, and reflecting on how interpretations evolve.

To address this challenge, we propose QualiVision, an interactive system for scaling qualitative video analysis. QualiVision aims to reduce persistent interaction frictions and supports reflective, iterative sensemaking practices in video-based qualitative analysis through two design incentives:

\begin{itemize}
    \item First, QualiVision will integrate video, transcripts, annotations, spreadsheets, and preliminary analytic reports within a synchronized workspace to reduce the workflow fragmentation. A spreadsheet-backed data model will maintain bidirectional synchronization between video evidence and structured analytic records, while parallel coding streams will allow researchers to preserve alternative interpretations and track how codes evolve over time.
    \item Second, QualiVision will use visualization and AI as heuristic aids for reflection, helping researchers notice emerging patterns earlier, compare relationships across coding streams at first glance, question their assumptions, and surface potentially overlooked insights.
\end{itemize}

\section{Background}
Prior work on video analysis has addressed several related but distinct problems. 
Video retrieval systems treat video primarily as a corpus to be indexed, queried, and retrieved efficiently~\cite{b2}. While useful when analytic targets are known in advance, this framing fits early-stage qualitative analysis less well, where researchers may not yet know which events, behaviors, or patterns will become meaningful. 
Other work seeks to support researchers' practical workflows around video data. 
Annotation tools such as ANVIL support structured, multi-layer annotation of audiovisual interaction~\cite{b3}, while systems such as ChronoViz~\cite{b4} help researchers synchronize video with other time-coded evidence streams. Many systems address adjacent challenges: Reflexis~\cite{b5} supports reflexivity, code evolution, and collaborative interpretation in qualitative analysis, uxSense~\cite{b6} explores automation for reducing coding labor by surfacing candidate patterns, and comparison visualization systems support the inspection of multimodal behavioral data~\cite{b7}. 
QualiVision builds on these efforts by asking how a complete qualitative video-analysis workflow can sustain reflexive, iterative sensemaking as video data scales. It explores how fragmented analytic activities can be integrated and where heuristic aids such as visualization and AI should fit within researchers' existing practices.

\begin{figure*}[t]
    \centering
    \includegraphics[width=0.95\textwidth]{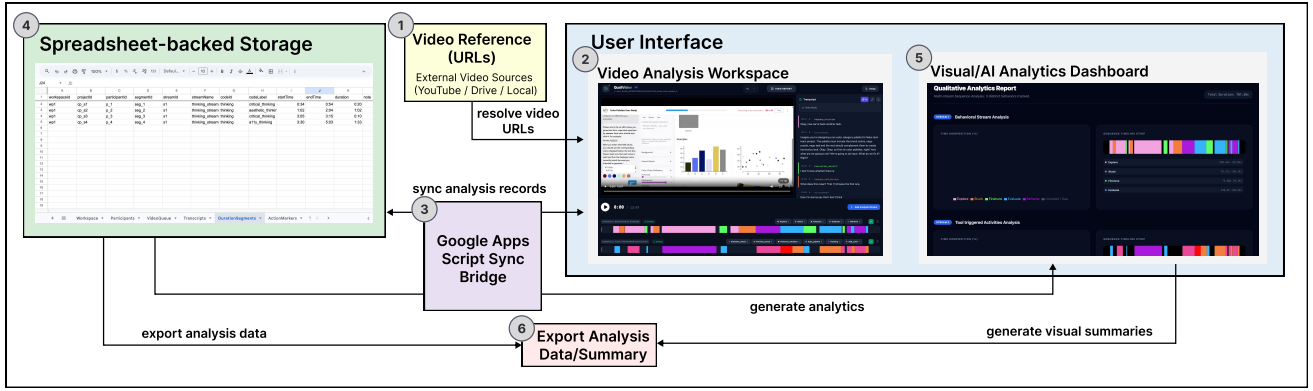}
    \caption{
QualiVision system architecture. \circnum{2} and \circnum{5} show the interfaces of the preliminary QualiVision prototype.
\circnum{1} External videos are referenced through URLs rather than hosted by QualiVision.
\circnum{2} The video analysis workspace supports video review, transcript inspection, coding, and note-taking.
\circnum{3} The Google Apps Script Sync Bridge synchronizes analysis records between the spreadsheet-backed storage and the video analysis interface.
\circnum{4} The spreadsheet-backed storage stores analysis data such as codes, transcripts, and memos.
\circnum{5} The Visual/AI analytics dashboard renders visual summaries and AI-assisted prompts.
\circnum{6} The export module supports exporting analysis data and visual summaries.
}
    \label{fig:qualivision}
\end{figure*}

\section{Preliminary Prototype and Future Work}

I have implemented an initial QualiVision prototype (Fig~\ref{fig:qualivision}) to explore how video-based qualitative analysis can be supported within a shared analytic workspace. The prototype brings together video playback, transcript editing, user-uploaded or AI-generated transcripts, multi-stream coding timelines, undo/redo support, simple visual summaries, local data packaging, and preliminary report generation. It has been used in my own video-based user study analyses to help coordinate video review, transcript inspection, and annotations across multiple coding streams.

The prototype is still under development and has not yet fully achieved its design goals. Many of these goals, including cross-stream comparison through visualization, spreadsheet-backed synchronization, and AI-assisted reflection, emerged through using the tool in practice. The next stage of the work is to refine and extend the existing implementation around these needs. To do so, two major challenges remain central.


The first challenge is technical. Moving from a local prototype to a spreadsheet-backed analysis environment requires a stable data model that can map project metadata, videos, transcripts, coded segments, coding streams, memos, visual summaries, and AI prompts while preserving their links to the video timeline. It also requires reliable bidirectional synchronization  between the interface and the spreadsheet backend, including edit batching, conflict detection, provenance tracking, and access control, so that compatibility with researchers' spreadsheet practices does not introduce data drift, broken links, or silent overwrites.

Second, QualiVision should not simply make visualization and AI available uncritically. Instead, it needs to determine which forms of support are useful for qualitative video sensemaking. This requires a closer study of common video-analysis practices: what patterns researchers look for, how they compare codes across participants or analytic stages, and where visual summaries may overstate the objectivity or importance of particular patterns. AI support must also remain a reflective layer rather than an interpretive authority, raising questions about the boundaries of AI intervention: when prompts should appear, what they should surface, and how they can support reflection without weakening researchers' control over interpretation.



\section{Proposed Evaluation}

To evaluate QualiVision, I plan to conduct a Comparative Structured Observation (CSO)~\cite{b8} study together with a small number of case studies. In the CSO study, participants will use both a conventional multi-tool workflow and QualiVision on comparable video-analysis tasks, such as open coding, identifying emerging patterns, and revisiting evidence to explain how an interpretation is grounded. I will collect measures such as task completion time, self-reported cognitive load, and usability scores to descriptively examine whether QualiVision outperforms the baseline. Think-aloud data and post-study interviews will further help assess how QualiVision supports reflexive and iterative sensemaking, while also identifying potential usability problems.

As short-term tasks may not capture how qualitative analysis develops over time, I will also conduct case studies in which researchers use QualiVision for a fuller analysis process, potentially with their own video data. These case studies will complement the CSO study by showing how the system supports longer-term practices in scaling qualitative video analysis, and help reduce the influence of novelty and learning effects.


\begin{thebibliography}{00}









\bibitem{b1} Reis, H. T., Charles M. J., ``Handbook of research methods in social and personality psychology,'' Cambridge University Press, 2000.

\bibitem{b2} Flickner, M., et al., ``Query by image and video content: The QBIC system,'' computer, 1995.

\bibitem{b3} Kipp, M., ``Anvil-a generic annotation tool for multimodal dialogue,'' Proc. Eurospeech, 2001.

\bibitem{b4} Fouse, A. S., ``Navigation of time-coded data,'' University of California, San Diego, 2013.

\bibitem{b5} Ye, R., et al., ``Reflexis: Supporting Reflexivity and Rigor in Collaborative Qualitative Analysis though Design for Deliberation,'' CHI, 2026.

\bibitem{b6} Batch, A., Ji, Y., Fan, M., Zhao, J., Elmqvist, N., ``uxSense: Supporting user experience analysis with visualization and computer vision,'' IEEE Transactions on Visualization and Computer Graphics, 2023.

\bibitem{b7} Blascheck, T., Beck, F., Baltes, S., Ertl, T., Weiskopf, D., ``Visual analysis and coding of data-rich user behavior,'' VAST, 2016.

\bibitem{b8} Mackay, W. E., McGrenere, J., ``Comparative Structured Observation,'' ACM Transactions on Computer-Human Interaction, 2025.







\end{thebibliography}
\end{document}